\documentclass[aps,prd,twocolumn,showpacs,superscriptaddress,
preprintnumbers,nofootinbib,reprint]{revtex4-1}
\pdfoutput=1
\usepackage[T1]{fontenc}
\usepackage{bm,amsmath,amssymb,color,bbold}
\usepackage[pdftex]{graphicx}
\usepackage[colorlinks=true, pdfstartview=FitV, linkcolor=blue, citecolor=blue, urlcolor=blue]{hyperref}

\DeclareMathOperator\diag{diag}

\DeclareMathOperator\tr{tr}
\DeclareMathOperator\Tr{Tr}

\newcommand{\der}{\partial}

\newcommand{\calO}{\mathcal{O}}
\newcommand{\dd}{\mathrm{d}}
\newcommand{\bep}{\begin{pmatrix}} 
\newcommand{\eep}{\end{pmatrix}}

\renewcommand{\O}{\text{O}}
\newcommand{\U}{\text{U}}
\newcommand{\1}{\mathbb{1}}

\renewcommand{\epsilon}{\varepsilon}

\def\ba#1\ea{\begin{align}#1\end{align}}
\def\akakko#1{\left\langle #1 \right\rangle}
\def\mkakko#1{\left( #1 \right)}
\def\ckakko#1{\left\{ #1 \right\}}
\def\kkakko#1{\left[ #1 \right]}
\def\kakko#1{\left\langle #1 \right\rangle}

\newcommand{\Rp}{R^{\phi}_k}
\newcommand{\Rs}{R^{\sigma}_k}
\newcommand{\dert}{\tilde{\der}_k}
\newcommand{\hatg}{\hat{g}}

\begin{document}
\preprint{RIKEN-QHP-219} \title{\boldmath Nonperturbative RG analysis of
five-dimensional $\O(N)$ models with cubic interactions}

\author{Kazuhiko Kamikado} \affiliation{Department of Physics, Tokyo
Institute of Technology, Meguro, Tokyo, 152-8551, Japan}
\author{Takuya Kanazawa} \affiliation{iTHES Research Group and Quantum
Hadron Physics Laboratory, RIKEN, Wako, Saitama 351-0198, Japan}
\allowdisplaybreaks
\begin{abstract}
 We reconsider critical properties of $\O(N)$ scalar models with cubic interactions in $d>4$ dimensions 
 using functional renormalization group equations. Working at next-to-leading order in the   
 derivative expansion, we find non-trivial IR fixed points at small and intermediate $N$ 
 from beta functions for relevant cubic terms. The putative fixed point at large $N$ 
 suggested recently by higher spin holography and the $\epsilon$-expansion is also discussed, 
 with an emphasis on stability of the effective potential.
\end{abstract}
\maketitle

\section{Introduction}
Theoretical understanding of critical phenomena and universality in the
framework of renormalization group is a milestone in modern theoretical
physics \cite{Goldenfeld:1992qy,ZJbook,Pelissetto:2000ek}.  Among others
the $\O(N)$ models have been thoroughly investigated with various
methods [e.g., the celebrated $\epsilon$-expansion, $1/N$ expansion,
high-temperature expansion, Monte Carlo simulation and the functional
renormalization group (FRG)] and nowadays there seems to be a solid
theoretical ground for our understanding of $\O(N)$-symmetric critical
points in $2<d<4$ dimensions. In the upper critical dimension $d=4$ the
Wilson-Fisher fixed point merges with the Gaussian fixed point and so
far no nontrivial stable fixed point with physically acceptable
properties has been found in $d\geq 4$, in agreement with the Ginzburg
criterion for a mean-field theory. In addition, there are even rigorous
proofs of triviality for $N=1$ and $2$
\cite{Frohlich:1982tw,Aizenman:1982ze}.  Recent works by Fei et
al.~\cite{Fei:2014yja,Fei:2014xta} suggested, however, possible
existence of a unitary non-Gaussian fixed point in $4<d<6$ at least for
sufficiently large $N$. Their argument is supported by higher-spin
AdS/CFT dualities \cite{Klebanov:2002ja,Giombi:2014iua}.  While the
original approach \cite{Fei:2014yja,Fei:2014xta} (see also
\cite{Gracey:2015tta,Herbut:2015zqa}) was based on the
$\epsilon$-expansion from $d=6$ dimensions, it would be quite desirable
to perform independent checks with nonperturbative methods. In this
regard, the conformal bootstrap approach
\cite{Nakayama:2014yia,Bae:2014hia,Chester:2014gqa} and FRG
\cite{Rosten:2008ts,Percacci:2014tfa,Mati:2014xma} 
have so far yielded contrasting results as to the existence of a
new fixed point: the former supports the claim of
\cite{Fei:2014yja,Fei:2014xta} whereas the latter does not.  In this
paper, we investigate the putative critical point in $d=5$ by means of
FRG, not based on the conventional $(\phi_i \phi_i)^2$-type formulation
with $N$ scalars but on the cubic $\O(N)$ model with $N+1$ scalars
\cite{Fei:2014yja}.  One of the advantages of FRG is that one can
directly work in $d=5$ with no need for dimensional continuation from
$d=4$ or $6$.  Our analysis suggests that no nontrivial critical point
exists at large $N$, in accordance with
\cite{Rosten:2008ts,Percacci:2014tfa,Mati:2014xma}.

This paper is structured as follows. In Sec.~\ref{sc:model} we define
the model, explain the effective average action approach and present the
flow equation at next-to-leading order in the derivative
expansion. The structure of the flow at finite $N$ is sketched.  In
Sec.~\ref{sc:largeN} the flow in the large-$N$ limit is discussed and
compared with the flow from $\epsilon$-expansion. We conclude in
Sec.~\ref{sc:concl}.  In \autoref{sc:appen} the derivation of the flow
equations is outlined.

\section{\boldmath\label{sc:model}RG equation for a cubic $\O(N)$ theory}
The functional renormalization group (FRG) is a powerful nonperturbative
method to solve problems with multiple scales in quantum field theory
and statistical physics.  In this approach, we consider the
scale-dependent effective average action, $\Gamma_k[\phi]$, which obeys
the exact flow equation
\cite{Nicoll:1977hi,Wetterich:1992yh,Morris:1993qb,Morris:1994ie} \ba
\der_k \Gamma_k & = \frac{1}{2}\,\mathrm{STr}
\ckakko{\big(\Gamma_k^{(2)} + R_k \big)^{-1}\der_k R_k}\,.
\label{eq:FRGeq} \ea Here STr and $R_k$ denotes a functional trace in
superspace and a regulator of the flow, respectively.
$\Gamma_{k=\Lambda}$ equals the classical action on the microscopic
scale, while $\Gamma_{k=0}$ is nothing but the full 1PI effective
action.  For more details, we refer the reader to
\cite{Bagnuls:2000ae,Berges:2000ew,
Delamotte:2003dw,Delamotte:2007pf,Rosten:2010vm,Braun:2011pp,
Wipf:2013vp,Nagy:2012ef}.

In this work we apply the FRG method to analyze the cubic $\O(N)$ model
\cite{Fei:2014yja}
\ba
	S & = \int \dd^dx \Big[
	\frac{1}{2}(\der_\mu\phi_i)^2 + 
	\frac{1}{2}(\der_\mu \sigma)^2 
	+ \frac{g_{1}}{2}\sigma \phi_i \phi_i + 
	\frac{g_{2}}{6}\sigma^3
	\Big]\,,  
	\label{eq:initcon}
\ea
which is perturbatively renormalizable in $d=6$.  The index $i$ runs
from 1 to $N$ and we leave $d$ and $N$ arbitrary at this stage. Despite
the precarious cubic potential, scalar theories like \eqref{eq:initcon}
have long been investigated due to their relevance to the Yang-Lee edge
singularity \cite{Fisher:1978pf,deAlcantaraBonfim:1981sy}, percolation
problems \cite{Fortuin:1971dw,deAlcantaraBonfim:1981sy} and more
recently, a six-dimensional generalization of the $a$-theorem
\cite{Grinstein:2014xba,Grinstein:2015ina} and $\mathcal{PT}$-symmetric
field theories \cite{Bender:2012ea,Bender:2013qp}.

Let us recall that, in the conventional setup of the $\O(N)$ vector
model, the coupling for $(\phi_i\phi_i)^2$ is marginal in $d=4$ and
irrelevant in $d>4$. Then a nontrivial fixed point, if any, should
appear as a UV fixed point \cite{Percacci:2014tfa,Mati:2014xma}. By
contrast, the action \eqref{eq:initcon} has two cubic couplings that are
marginal in $d=6$, so we may look for a nontrivial IR fixed point in
$d<6$. This situation is reminiscent of the Wilson-Fisher fixed point in
$d=3$ which can be identified either from a nonlinear sigma model in
$d=2+\epsilon$ as a UV fixed point, or from a quartic scalar theory in
$d=4-\epsilon$ as an IR fixed point \cite{ZJbook}.

While the flow equation \eqref{eq:FRGeq} itself is exact, one needs to
project it to a finite-dimensional functional space to make practical
calculations feasible. In this work, we employ the Ansatz
\ba
	\Gamma_{k}[\phi,\sigma] & = \int \dd^dx \left[
	\frac{Y_{k}}{2}(\der_\mu\phi_i)^2  
	+ \frac{Z_{k}}{2}(\der_\mu \sigma)^2
	+ U_k(\rho,\sigma)
	\right]  
	\label{eq:ansatz}
\ea
with $\displaystyle \rho\equiv \frac{1}{2}\phi_i \phi_i$ for the
truncated effective action at the scale $k$.  We could also add a term
$(\der_\mu \rho)^2$ that contributes to the difference of anomalous
dimensions for the radial mode and the Nambu-Goldstone modes, but here
it is omitted due to its high canonical dimension.  The factors $Y_k$
and $Z_k$ are the wave function renormalization for $\phi_i$ and
$\sigma$, respectively, and $U_k$ is the running effective
potential. The approximation for \eqref{eq:ansatz}, which is at the
next-to-leading order in the derivative expansion, is called the
improved local potential approximation (LPA$'$); when $Y_k=Z_k\equiv 1$
and only $U_k$ is running, this is the leading order in the derivative
expansion and is called LPA.  Both LPA and LPA$'$ have been successful
in describing various critical phenomena
\cite{Tetradis:1993ts,Bagnuls:2000ae,Berges:2000ew,
Delamotte:2003dw,Delamotte:2007pf,Nagy:2012ef}.  Although there is no
small parameter that controls the expansion of the truncated effective
action, it is known that LPA and LPA$'$ work better when the anomalous
dimension of fields is numerically small.  In the large-$N$ limit of the
quartic $\O(N)$ vector model where the anomalous dimension vanishes, LPA
becomes \emph{exact} for the effective potential
\cite{D'Attanasio:1997he,ZJbook,Moshe:2003xn}.

In this work we employ the optimized regulator devised by Litim
\cite{Litim:2000ci,Litim:2001up}
\ba
	\begin{Bmatrix} \Rp(p) \\ \Rs(p) \end{Bmatrix} 
	& = \begin{Bmatrix} Y_{k} \\ Z_{k} \end{Bmatrix}
	\times (k^2-p^2) \Theta(k^2-p^2) 
	\quad \text{for}~\begin{Bmatrix} \phi_i 
	\\ \sigma \end{Bmatrix},
	\label{eq:regR}
\ea
where $\Theta$ is the Heaviside step function.  The flow equations for
$Y_k$, $Z_k$ and $U_k$ can now be obtained straightforwardly by plugging
\eqref{eq:ansatz} and \eqref{eq:regR} into \eqref{eq:FRGeq}. Full
details of the derivation are presented in \autoref{sc:appen}.
Introducing the logarithmic scale $t \equiv \log(k/\Lambda)$, we obtain
\begin{widetext}
	\ba
		\der_t U_k & = \mu_d k^{d+2}\Bigg[
		\frac{
			Y_k k^2 + \frac{\der U_k}{\der \rho} 
			+ 2\rho \frac{\der^2 U_k}{\der \rho^2} 
		}
		{
			(Z_k k^2 + \frac{\der^2 U_k}{\der\sigma^2})
			(Y_k k^2 + \frac{\der U_k}{\der \rho} 
			+ 2\rho \frac{\der^2 U_k}{\der \rho^2})
			- 2\rho (\frac{\der^2 U_k}{\der\rho \der \sigma})^2
		}
		\mkakko{ 1-\frac{\eta_{\sigma}}{d+2} } Z_k 
		\notag 
		\\
		& \qquad \qquad  + \Bigg\{ 
		\frac{
			Z_k k^2 + \frac{\der^2 U_k}{\der\sigma^2} 
		}
		{
			(Z_k k^2 + \frac{\der^2 U_k}{\der\sigma^2})
			(Y_k k^2 + \frac{\der U_k}{\der \rho} 
			+ 2\rho \frac{\der^2 U_k}{\der \rho^2})
			- 2\rho (\frac{\der^2 U_k}{\der\rho \der \sigma})^2
		}
		+ \frac{ N-1 }{Y_k k^2 + \frac{\der U_k}{\der \rho}}
		\Bigg\} 
		\mkakko{ 1-\frac{\eta_{\phi}}{d+2} }Y_k 
		\Bigg] \,,  
		\label{eq:uff}
		\\
		\eta_\phi & \equiv -\der_t \log Y_k 	
		\\
		& =  2 \mu_d k^{d+2}\kakko{\frac{\der^2 U_k}{\der \sigma\der \rho}}^2  
		\frac{Z_k}{
			\mkakko{Y_k k^2 + \kakko{\frac{\der U_k}{\der \rho}}}^2 
			\mkakko{Z_k k^2 + \kakko{\frac{\der^2 U_k}{\der \sigma^2}}}^2
		}\,,
		\label{eq:ephff}
		\\
		\eta_\sigma & \equiv -\der_t \log Z_k 
		\\
		& = \mu_d k^{d+2} \kakko{\frac{\der^3 U_k}{\der \sigma^3}}^2  
		\frac{Z_k}{
			\mkakko{Z_k k^2 + \kakko{\frac{\der^2 U_k}{\der \sigma^2}}}^4  
		}
		+ N \mu_d k^{d+2} \kakko{\frac{\der^2 U_k}{\der \sigma\der \rho}}^2  
		\frac{Y_k^2/Z_k}{
		\mkakko{Y_k k^2 + \kakko{\frac{\der U_k}{\der \rho}}}^4 
		}\,, 
		\label{eq:etff}
	\ea
\end{widetext}
where 
\ba
	\mu_d \equiv 
	\frac{1}{(4\pi)^{d/2}\Gamma\big(\frac{d}{2}+1\big)}
	\label{eq:mudefi}
\ea
and the bracket $\kakko{\cdots}$ denotes the value in a fixed background
$(\sigma(x),\phi_i(x))=(\sigma_0, \vec{0})$. Although the anomalous
dimensions in LPA$'$ are sometimes evaluated at the running minimum of
the effective potential for better convergence
\cite{Tetradis:1993ts,Aoki:1998um}, here we set $\phi_i=\vec{0}$ for a
technical reason and leave $\sigma_0$ arbitrary at this stage.

To investigate the scaling behavior near the fixed point it is
convenient to make all variables dimensionless by proper powers of
$k$. We thus define
\begin{subequations}
	\ba
		u_t(r,s) & = k^{-d} U_k(\rho,\sigma) \,,
		\\
		r & = k^{2-d} Y_k \rho\,,
		\\ 
		s & = k^{\frac{2-d}{2}} \sqrt{Z_k}\, \sigma\,,  
	\ea
\end{subequations}
which leads to the dimensionless flow equations
\begin{widetext}
\ba
	& \der_t u_t + du_t + (2-d-\eta_{\phi}) r \der_r u_t + 
	\frac{1}{2} (2-d-\eta_\sigma)  s \der_s u_t  
	\notag
	\\
	= ~& \mu_d \Bigg[
	\frac{
		1 + \der_r u_t + 2r \der^2_r u_t
	}
	{
		(1 + \der_s^2 u_t )(1 + \der_r u_t + 2 r \der^2_r u_t)
		- 2 r (\der_r \der_s u_t)^2
	}
	\mkakko{ 1-\frac{\eta_{\sigma}}{d+2} } 
	\notag 
	\\
	& \qquad ~ + \Bigg\{ 
	\frac{1 + \der_s^2 u_t }
	{
		(1 + \der_s^2 u_t )(1 + \der_r u_t + 2 r \der^2_r u_t)
		- 2 r (\der_r\der_s u_t)^2 
	}
	+ 
	\frac{ N-1 }{1 + \der_r u_t }
	\Bigg\} \mkakko{ 1-\frac{\eta_{\phi}}{d+2} }
	\Bigg] \,,
	\label{eq:dimlessflow}
\ea
\end{widetext}
\ba
	\eta_\phi & = 2 \mu_d \frac{\kakko{\der_r\der_s u_t}^2}
	{
		(1+\kakko{\der_r u_t})^2 (1+\kakko{\der_s^2 u_t})^2
	} \,,
	\label{eq:dimlessfloweta1}
	\\
	\eta_\sigma & = \mu_d  \left[
	\frac{\kakko{\der_s^3 u_t}^2}{
		\mkakko{  1 + \kakko{\der_s^2 u_t}  }^4  
	}
	+ N \frac{\kakko{\der_r \der_s u_t}^2}{
		\mkakko{1 + \kakko{\der_r u_t}}^4 
	}
	\right] .
	\label{eq:dimlessfloweta2}
\ea
As a small check, notice that if we neglect the $s$-dependence of $u_t$,
we find $\eta_\phi=\eta_\sigma=0$ and recover the flow equation for
$u_t$ in the quartic $\O(N)$ model with no $\sigma$ field
\cite{Litim:2001up}.%
\footnote{There remains an irrelevant constant in
the RHS of \eqref{eq:dimlessflow}, which represents the contribution of
the free massless scalar $\sigma$.}

\begin{figure}[b]
	\centering
	\includegraphics[width=60mm]{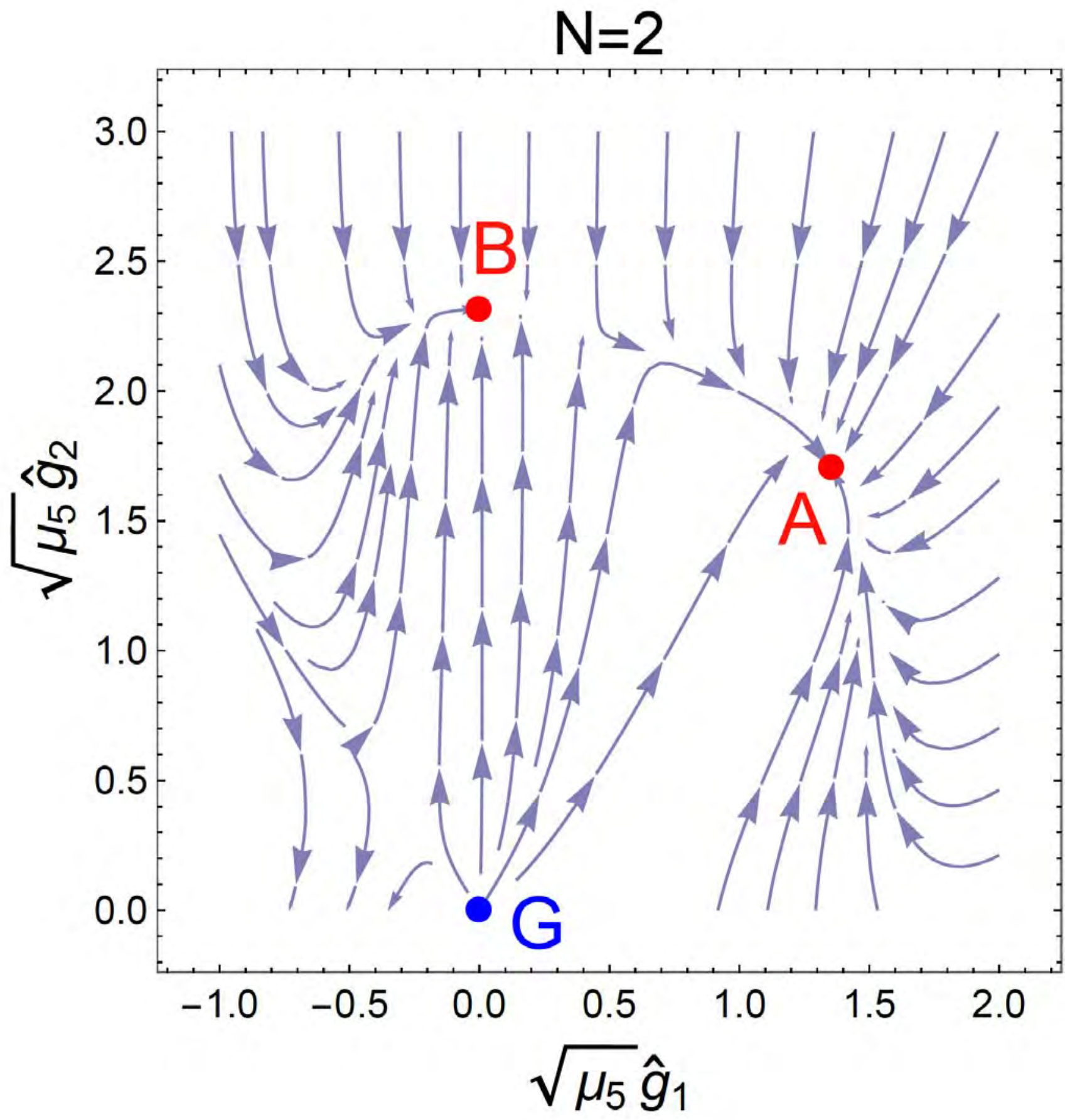}
	\caption{\label{fg:N=2flow}RG flow towards IR with $N=2$ and $d=5$ 
	for the minimally truncated Ansatz \eqref{eq:uthree}. The flow diagram is symmetric 
	under $(\hatg_1,\hatg_2)\leftrightarrow(-\hatg_1,-\hatg_2)$ and only the region 
	with $\hatg_2\geq 0$ is shown. The blobs $A$ and $B$ represent 
	IR-stable fixed points, while the blob $G$ is the Gaussian fixed point which is unstable in IR. There are two more unstable fixed points 
	in the figure (not shown).
	}
\end{figure}

To make a comparison with the $\epsilon$-expansion around $d=6$, let us
substitute a simplistic Ansatz [cf.~\eqref{eq:initcon}]
\ba
	u_t(r,s) & = \hatg_1(t) rs + \frac{\hatg_2(t)}{6}s^3 
	\label{eq:uthree}
\ea
into \eqref{eq:dimlessflow} and expand the RHS in powers of $r$ and
$s$. Furthermore we choose to evaluate $\akakko{\dots}$ in
\eqref{eq:dimlessfloweta1} and \eqref{eq:dimlessfloweta2} at
$\sigma_0=0$. This yields the beta functions of $\hatg_1$ and $\hatg_2$,
\\
\scalebox{0.93}{
\begin{minipage}{.5\textwidth}
\begin{subequations}
\ba
	\frac{\dd \hatg_1}{\dd t} & = - \frac{6 - d - 2\eta_\phi - \eta_\sigma}{2}\hatg_1 
	\notag
	\\
	& \quad -2 \mu_d \hatg_1^2 \kkakko{
		\mkakko{ 3-\frac{2\eta_\phi + \eta_{\sigma}}{d+2} } \hatg_1 
		+ \mkakko{ 3-\frac{\eta_\phi + 2\eta_{\sigma}}{d+2} } \hatg_2
	},
	\\
	\frac{\dd \hatg_2}{\dd t} & = - \frac{6 - d - 3\eta_\sigma}{2}\hatg_2  
	\notag
	\\
	& \quad - 6 \mu_d \kkakko{
		\mkakko{ 1-\frac{\eta_{\phi}}{d+2} }N\hatg_1^3 
		+ 
		\mkakko{ 1-\frac{\eta_{\sigma}}{d+2} }\hatg_2^3
	}, \hspace{-5pt}
\ea
\label{eq:bexpa}%
\end{subequations}
\end{minipage}
}
\\
and 
\ba
	\eta_\phi = 2\mu_d \hatg_1^2\,, 
	\quad 
	\quad 
	\eta_\sigma & = \mu_d ( N \hatg_1^2+\hatg_2^2 )\,.
	\label{eq:etapt}
\ea
If we suppress $\eta_\phi$ and $\eta_\sigma$ in the square brackets of
\eqref{eq:bexpa}, insert $d=6-\epsilon$ in the first term of
\eqref{eq:bexpa} and set $\mu_d=\mu_6 =
\frac{1}{(4\pi)^{d/2}\Gamma(\frac{d}{2}+1)}\Big|_{d=6}=\frac{1}{6(4\pi)^3}$,
then $\dd \hatg_1/\dd t$ and $\dd \hatg_2/\dd t$ exactly match the beta
functions from the $\epsilon$-expansion at one loop \cite{Fei:2014yja}.
Also $\eta_\phi/2$ and $\eta_\sigma/2$ in \eqref{eq:etapt} exactly match
the anomalous dimensions of $\phi$ and $\sigma$ in \cite{Fei:2014yja}.%
\footnote{Coincidence of the perturbatively expanded FRG and the
one-loop $\epsilon$-expansion has been reported for a $\U(2)\times\U(2)$
scalar model \cite{Fukushima:2010ji}.}  However, one difference from the
$\epsilon$-expansion is that $\eta_\phi$ and $\eta_\sigma$ multiply
$\hatg_1^3$, $\hatg_1^2\hatg_2$ and $\hatg_2^3$ in
\eqref{eq:bexpa}. This reflects the fact that higher-order contributions
are incorporated differently in the loop expansion and FRG.  In
Fig.~\ref{fg:N=2flow} we display the flow diagram of \eqref{eq:bexpa}
for $d=5$ and $N=2$.
Intriguingly, besides the Gaussian fixed point at the origin $(0,0)$,
there are two nontrivial IR-stable fixed points $A$ and $B$ that are
absent in the $\epsilon$-expansion at one loop.  The anomalous
dimensions at these fixed points are
\ba
	\begin{split}
		A:&~(\eta_\phi,\eta_\sigma) = (3.67,~~6.58)\,,
		\\
		B:&~(\eta_\phi,\eta_\sigma) = (0,~~5.36)\,,
	\end{split}
	\label{eq:ABeta}
\ea
respectively.  The point $B$ is present for any $N\geq 2$, whereas the
point $A$ disappears for $N\geq 19$. The large-$N$ fixed point taken up
in \cite{Fei:2014yja,Fei:2014xta} comes into existence only for
$N\gtrsim 820$, which is slightly below the threshold $\sim 1039$ in the
one-loop $\epsilon$-expansion \cite{Fei:2014yja}.  The question one must
ask is whether $A$ and $B$ do represent physical critical points or
not. In this regard we have to recognize that the anomalous dimensions
\eqref{eq:ABeta} at $A$ and $B$ are dangerously large and threaten the
validity of LPA$'$.  Note also that the presence of \emph{multiple}
IR-stable fixed points leads to a bewildering consequence that two
systems in the same dimension and sharing the same symmetry may exhibit
different universal behaviors without fine-tuning, depending on which
basin of attraction the initial parameters fall in. This exotic
situation is not expected to arise in physically sound systems. Based on
these observations, we would like to take a conservative point of view
that the fixed points $A$ and $B$ are artifacts of the truncation
\eqref{eq:uthree}.%
\footnote{IR fixed points at small $N$ were also
reported in \cite{Herbut:2015zqa} within $\O(N)$ models with tensorial
interaction. Whether our fixed points $A$ and $B$ have anything to do
with \cite{Herbut:2015zqa} is unclear.}  One way to test this idea would
be by extending the truncation of $u_t$ to higher orders and check
stability of $A$ and $B$, taking carefully into account a number of
subtleties associated with the polynomial truncation method
\cite{Morris:1994ki,Aoki:1998um, Grahl:2014fna,Mati:2014xma}.

\section{\boldmath\label{sc:largeN}Large $N$}
Next we turn to the analysis at $N\gg 1$. In this limit the flow
equations are simplified. For the counting \ba u_t(r,s) \sim r \sim N
\quad \text{and} \quad s \sim \sqrt{N}\,, \ea one obtains, at leading
order, \ba & \hspace{-12pt} \der_t u_t + du_t + (2-d) r \der_r u_t +
\frac{1}{2} (2-d-\eta_\sigma) s \der_s u_t \notag \\ & = \mu_d
\frac{N}{1+\der_r u_t}\,, \label{eq:largeNflow1} \\ \eta_{\phi} & =
\calO(1/N)\,, \\ \eta_\sigma & = \mu_d N \frac{\kakko{\der_r \der_s
u_t}^2}{ \mkakko{1 + \kakko{\der_r u_t}}^4 } = \calO(1)\,.
\label{eq:etaN} \ea In this model $\eta_\sigma$ does \emph{not} vanish
in the large-$N$ limit, in contrast to quartic $\O(N)$ vector models
where the anomalous dimensions of scalars vanish in the same limit
\cite{ZJbook}.%
\footnote{In fermionic theories, the nonvanishing $\eta$
of scalars in the many-flavor limit is well known
\cite{ZinnJustin:1991yn,Braun:2010tt}.}  Equation \eqref{eq:largeNflow1}
implies that the effective potential $u_\star$ at the RG fixed point,
called \emph{a scaling solution}, must satisfy \ba du_{\star} + (2-d) r
\der_r u_{\star} + \frac{1}{2} (2-d-\eta_\sigma) s \der_s u_{\star} =
\mu_d \frac{N}{1+\der_r u_{\star}} \,.  \label{eq:largeNflow2} \ea
Evidently there is a trivial solution $u_\star=\mu_d N/d$ corresponding
to the Gaussian fixed point for any $N$ and $d$.  Whether a globally
well-defined nontrivial scaling solution to \eqref{eq:largeNflow2} [to
be solved self-consistently with \eqref{eq:etaN}] exists or not in $d=5$
is our central concern here.  The advantage of this FRG approach as
compared to the $\epsilon$-expansion is that one can search for a fixed
point directly in $d=5$ \emph{without} placing any specific Ansatz for
the effective potential. That said, it is usually hard to solve a
fixed-point equation like \eqref{eq:largeNflow2} analytically.  One may
resort to solving it numerically, by integrating the partial
differential equation starting from the origin.  This method reveals
that most of numerical solutions thus obtained encounter a singularity
at a finite value of the field, as emphasized by Morris
\cite{Morris:1994ki,Morris:1996nx,Morris:1998da} (see also
\cite{Codello:2012sc,Hellwig:2015woa}). Even when the flow could be
smoothly integrated over the entire field values, it may not be
necessarily bounded from below. In these cases one has to conclude that
there is no physical critical point. The main message here is that
analyzing a truncated effective potential just around the origin is
fallacious since it masks pathological global properties of the
potential \cite{Percacci:2014tfa}.

Now, coming back to \eqref{eq:largeNflow2}, 
one finds that in the limit $s\to 0$, 
\ba
	du_{\star}(r,0) + (2-d) r \der_r u_{\star}(r,0) 
	= \mu_d \frac{N}{1+\der_r u_{\star}(r,0)} \,, 
	\label{eq:ffflll}
\ea
which coincides exactly with the fixed-point equation 
with the optimized regulator 
for the quartic $\O(N)$ model with no $\sigma$ field \cite{Litim:2001up}. 
The structure of solutions to \eqref{eq:ffflll} for $d>4$ has already been 
thoroughly investigated in \cite{Percacci:2014tfa,Mati:2014xma} 
with the conclusion that they are either unbounded from below, 
or beset with singularities at a finite field.%
\footnote{The lack of a lower bound for the potential is consistent with 
the fact that the UV fixed point value for the coupling $(\phi_i\phi_i)^2$ 
in the quartic $\O(N)$ model with no $\sigma$ is \emph{negative} 
in $d=4+\epsilon$, corresponding to a bottomless potential \cite{ZJbook,
Giombi:2014iua}.}
The fact that $u_{\star}(r,0)$ 
is pathological forces us to conclude that \eqref{eq:largeNflow2} 
for $d>4$ possesses no acceptable scaling solution other than the trivial one.  
Note that the scalar $\sigma$ plays no role here, although  
the cubic potential of $\sigma$ in \eqref{eq:initcon} appears 
at first sight to be the major source of instability in this model.  
The conclusion above may not come as a total surprise 
if we make the following observation: 
the scaling dimension of $\sigma$ is equal to 2 at large $N$ 
\cite{Fei:2014yja} so that both $(\der \sigma)^2$ and $\sigma^3$ 
are irrelevant in $d=5$ and can be dropped without affecting 
physics in IR. Then $\sigma$ with the action 
$\sim \phi_i\phi_i\sigma+\sigma^2$ can be integrated out, 
thus recovering the ordinary $\O(N)$ vector model with 
$(\phi_i\phi_i)^2$ coupling.%
\footnote{An analogous argument shows the equivalence between 
the Gross-Neveu model and the Yukawa model in $2<d<4$ 
at large $N$ \cite{Hasenfratz:1991it,ZinnJustin:1991yn}.} 
Since the latter model does not possess a healthy scaling solution 
in $d>4$, the cubic model \eqref{eq:initcon} does not either.

\section{\label{sc:concl}Discussion}

In this paper, we have investigated a scalar $\O(N)$ model with cubic
interactions using the functional renormalization group (FRG) method at
next-to-leading order in the derivative expansion, for the purpose of
testing a recent conjecture \cite{Fei:2014yja} (backed up by higher-spin
AdS/CFT dualities \cite{Klebanov:2002ja,Giombi:2014iua}) that there is
an interacting unitary $\O(N)$-symmetric CFT in $d=5$ dimensions.  The
first analysis \cite{Fei:2014yja} based on the one-loop
$\epsilon$-expansion has already been extended to three
\cite{Fei:2014xta} and even four loops \cite{Gracey:2015tta}, confirming
that the cubic $\O(N)$ model has a non-Gaussian IR-stable fixed point in
$d=6-\epsilon$ dimensions if $N$ is above a certain threshold ($\lesssim
1000$). If true, this would herald new physics, defying the conventional
wisdom that scalar theories in $d\geq 4$ are trivial in the continuum
limit. In order to place this claim on firmer ground one needs to rely
on a nonperturbative approach.  Preceding FRG analyses
\cite{Percacci:2014tfa,Mati:2014xma} based on quartic $\O(N)$ vector
models with $N$ scalars have reported negative evidence as to the
existence of a nontrivial stable fixed point, in harmony with earlier
work \cite{Rosten:2008ts}. On the other hand, in this work, we start
directly from the cubic $\O(N)$ model considered in
\cite{Fei:2014yja,Fei:2014xta}. We found that there is no IR-stable
fixed point at large $N$, which corroborates
\cite{Percacci:2014tfa,Mati:2014xma}. It is worth mentioning that we did
not rely on a dimensional expansion from $d=4$ or $6$ but directly
worked in $d=5$, and made no specific Ansatz for the effective potential
to reach the above conclusion at large $N$.  Thus we are led to conclude
that the addition of a scalar $\sigma$ with cubic interactions does not
bring about qualitative differences from the quartic $\O(N)$ model. In
this regard we disagree with the conformal bootstrap approach
\cite{Nakayama:2014yia,Bae:2014hia,Chester:2014gqa} which seems to be in
favor of the putative fixed point.

Of course the analysis presented here is not completely free from
approximations; we have used a truncated action at next-to-leading order
in the derivative expansion.  However, the anomalous dimension of
$\sigma~(\sim 1/2)$ is not so large as to invalidate the derivative
expansion qualitatively.  We wish to also mention that FRG at this level
of approximation has been successful in many other circumstances
\cite{Morris:1993qb,Bagnuls:2000ae,Berges:2000ew,Delamotte:2007pf,Nagy:2012ef}.
If it transpires that the non-Gaussian fixed point does indeed exist,
but is invisible in FRG, then it is an imperative task to understand why
FRG fails to capture it. A deeper understanding of potential
deficiencies of FRG would be instrumental in identifying the origin of
discrepancy between FRG and other methods in fields such as QCD
with two flavors \cite{Pisarski:1983ms,
Fukushima:2010ji,Aoki:2012yj,Pelissetto:2013hqa,Grahl:2014fna,
Nakayama:2014sba,Kanazawa:2015xna,2016PhRvB..93f4405D}
where the nature of the chiral transition is still under debate, and frustrated magnets
\cite{1998JPCM...10.4707K,Delamotte:2003dw,Butti:2003nu,Calabrese:2004nt,Nakayama:2014sba}
where the existence of IR fixed points is disputed. On the other
hand, if the claimed $\O(N)$ critical theory is a non-unitary
metastable theory (as is indicated by
\cite{Percacci:2014tfa,Mati:2014xma} and this work), then a natural
question to ask is how to distinguish such illusionary fixed points from
physical ones within the conformal bootstrap approach.  In either
scenario our understanding of field theories can be deepened through a
further investigation on this issue.

Last but not least, the (non)existence of stable fixed points at finite
or small $N$ in $d>4$ is also of interest.  In this work, we have found
two IR-stable non-Gaussian fixed points in the real-coupling region for
$N=2$ and $d=5$.  This should not be taken at face value, however, given
the large anomalous dimensions at these points which are likely to be a
signal of the breakdown of the derivative expansion.  While we have not
attempted to explore the domain of imaginary couplings, the latter has
physical importance with regards to e.g., the Yang-Lee edge singularity
\cite{Fisher:1978pf}, percolation problems \cite{Fortuin:1971dw} and
$\mathcal{PT}$-symmetry \cite{Bender:2012ea}. For these applications our
flow equation \eqref{eq:uff} provides a useful point of departure for a
nonperturbative analysis in the future.

\vspace{0.5cm}
\noindent\textbf{Note added}
\vspace{0.2cm}\\
While this paper was at the final stage of preparation, we became aware of independent
work \cite{Eichhorn:2016hdi} where the same model was analyzed. 

\begin{acknowledgments}
 T.~K. was supported by the RIKEN iTHES project. 
\end{acknowledgments}

\appendix*
\section{\label{sc:appen}Flow equation}

In this appendix we derive the flow equations 
\eqref{eq:uff}, \eqref{eq:ephff} and \eqref{eq:etff} 
for the effective potential $U_k$ and the wave function renormalization 
$Y_k$ and $Z_k$. 

\subsection{\boldmath Flow of $U_k$} 

In a homogeneous background, \eqref{eq:FRGeq} may be evaluated 
in a plane wave basis: 
\ba
	\der_k U_k & = \frac{1}{2} \int \!\! \frac{\dd^d p}{(2\pi)^d} \tr 
	\left[ \frac{1}{\Gamma_k^{(2)} + R_k}\der_k R_k \right] \,,  
	\label{eq:U_flow}
\ea
with $R_k = \diag(\Rs, \Rp \1_N)$, see \eqref{eq:regR}.   
Note that both $\Gamma^{(2)}_k$ and $R_k$ are $(N+1)\times (N+1)$ 
matrices in the space of field components.  
Without loss of generality, any $\vec{\phi}$ can be rotated to the first direction 
as $\vec{\phi}=(\phi, 0, \dots, 0)$ so that 
$\rho\equiv \vec{\phi}^{\,2}/2 = \phi^2/2$.  
Then 
\begin{widetext}
	\ba
		\Gamma_k^{(2)} + R_k & = 
		\begin{pmatrix}
			Z_k p^2+ \Rs + \frac{\der^2 U_k}{\der\sigma^2} & 
			\phi \frac{\der^2 U_k}{\der\rho \der \sigma} & 0
			\\
			\phi \frac{\der^2 U_k}{\der\rho \der \sigma} & 
			Y_k p^2 + \Rp 
			+ \frac{\der U_k}{\der \rho} 
			+ \phi^2 \frac{\der^2 U_k}{\der \rho^2}  & 0 
			\\
			0 & 0 & (Y_k p^2 + \Rp + \frac{\der U_k}{\der \rho})\1_{N-1}  
		\end{pmatrix}\,.
	\ea
	Plugging this into \eqref{eq:U_flow}, one finds
	\ba
		\der_k U_k & = \frac{1}{2} \int \!\! \frac{\dd^d p}{(2\pi)^d} 
		\left[
			X_{\sigma} \der_k \Rs + X_{\phi}\der_k \Rp 
			+ (N-1) \frac{\der_k \Rp}{Y_k p^2 + \Rp 
			+ \frac{\der U_k}{\der \rho}}
		\right] 
		\label{eq:U_flow2}
	\ea
	with 
	\begin{subequations}
	\ba
		X_{\sigma} & \equiv 
		\frac{Y_k p^2 + \Rp + \frac{\der U_k}{\der \rho} 
		+ \phi^2 \frac{\der^2 U_k}{\der \rho^2}}
		{(Z_k p^2+ \Rs + \frac{\der^2 U_k}{\der\sigma^2})
		(Y_k p^2 + \Rp + \frac{\der U_k}{\der \rho} 
		+ \phi^2 \frac{\der^2 U_k}{\der \rho^2})
		- \phi^2 (\frac{\der^2 U_k}{\der\rho \der \sigma})^2}\,,
		\\
		X_{\phi} & \equiv 
		\frac{Z_k p^2+ \Rs + \frac{\der^2 U_k}{\der\sigma^2}}
		{(Z_k p^2+ \Rs + \frac{\der^2 U_k}{\der\sigma^2})
		(Y_k p^2 + \Rp + \frac{\der U_k}{\der \rho} 
		+ \phi^2 \frac{\der^2 U_k}{\der \rho^2})
		- \phi^2 (\frac{\der^2 U_k}{\der\rho \der \sigma})^2}\,. 
	\ea
	\end{subequations}
	This rather complicated appearance is caused by the mixing between $\rho$ and $\sigma$. 
	A simplification comes from the observation that 
	the presence of $\der_k \Rs$ and $\der_k \Rp$ in \eqref{eq:U_flow2} allows us to replace 
	$\Rs$ and $\Rp$ in $X_{\sigma,\phi}$ by $Z_k(k^2-p^2)$ and $Y_k (k^2-p^2)$, 
	respectively. This leads to the expression
	\ba
		\der_k U_k & = \frac{1}{2} \int \!\! \frac{\dd^d p}{(2\pi)^d} 
		\Bigg[
		\frac{
			(Y_k k^2 + \frac{\der U_k}{\der \rho} 
			+ 2\rho \frac{\der^2 U_k}{\der \rho^2}) 
			\der_k \Rs
		}
		{
			(Z_k k^2 + \frac{\der^2 U_k}{\der\sigma^2})
			(Y_k k^2 + \frac{\der U_k}{\der \rho} 
			+ 2\rho \frac{\der^2 U_k}{\der \rho^2})
			- 2\rho (\frac{\der^2 U_k}{\der\rho \der \sigma})^2
		}
		\notag 
		\\
		& \qquad + \frac{
			(Z_k k^2 + \frac{\der^2 U_k}{\der\sigma^2})\der_k \Rp 
		}
		{
			(Z_k k^2 + \frac{\der^2 U_k}{\der\sigma^2})
			(Y_k k^2 + \frac{\der U_k}{\der \rho} 
			+ 2\rho \frac{\der^2 U_k}{\der \rho^2})
			- 2\rho (\frac{\der^2 U_k}{\der\rho \der \sigma})^2
		}
		+ (N-1) \frac{\der_k \Rp}{Y_k k^2 + \frac{\der U_k}{\der \rho}}
		\Bigg]\,. 
		\label{eq:U_flow3}
	\ea
\end{widetext}
In the above we replaced $\phi^2$ by $2\rho$. 
Now the remaining integral over $p$ 
can be easily done with the formulas
\begin{subequations}
	\ba
		\int \!\! \frac{\dd^d p}{(2\pi)^d} \der_k \Rs  
		& = 2 \mu_d k^{d+1}
		\mkakko{ \frac{1}{d+2} k  \der_k Z_k  + Z_k } \,,
		\\
		\int \!\! \frac{\dd^d p}{(2\pi)^d} \der_k \Rp  
		& = 2 \mu_d k^{d+1}
		\mkakko{ \frac{1}{d+2} k  \der_k Y_k  + Y_k }   \,,
	\ea
\end{subequations}
with $\mu_d$ defined in \eqref{eq:mudefi}, 
which finally yields the flow equation \eqref{eq:uff} for $U_k$.

\subsection{\boldmath Flow of $Y_k$ and $Z_k$}

Let us evaluate \eqref{eq:FRGeq} in an inhomogeneous background 
\ba
	(\sigma \big|\, \phi_1,\dots,\phi_N) = 
	\mkakko{ \sigma_0 + u(x) \big|\, t(x),0,\dots,0 }  
	\label{eq:inhbg}
\ea
for which $\rho \equiv {\vec{\phi}}^{\,2}/2= t^2/2$. 
Then the matrix elements of $\Gamma_k^{(2)}$ admit an expansion 
in powers of $u$ and $t$ around $\sigma_0$. For instance
\ba
	\frac{\der^2 U_k}{\der \sigma\der \rho} 
	& =
	\kakko{\frac{\der^2 U_k}{\der \sigma\der \rho} } + 
	\kakko{\frac{\der^3 U_k}{\der \sigma^2 \der \rho}} u 
	+ \frac{1}{2}\kakko{\frac{\der^4 U_k}{\der \sigma^3 \der \rho}} u^2 
	\notag
	\\
	& \quad 
	+ \frac{1}{2}\kakko{\frac{\der^3 U_k}{\der \sigma\der \rho^2}} t^2 
	+\calO(u^3, ut^2)\,, 
\ea
where $\kakko{\dots}$ is to be evaluated at $(\sigma,\vec{\phi})=(\sigma_0,\vec{0})$ 
and we exploited the fact that terms odd in $t$ do not show up in the expansion. 
This way we obtain
\ba
	\Gamma_k^{(2)} + R_k & = A_0 + A_1 + A_2 + 
	{\cal O}(u^3, u^2 t, ut^2, t^3)\,, 
\ea
where $A_{0,1,2}$ are $(N+1)\times(N+1)$ matrices in the field space, 
defined as 
\ba
	& A_0 \equiv 
	\notag
	\\
	& \left(\begin{array}{cc} 
		- Z_k \der^2 + \Rs + \big\langle \frac{\der^2 U_k}{\der \sigma^2} \big\rangle & 0 
		\\
		0 & \big[ - Y_k \der^2 + \Rp + \big\langle \frac{\der U_k}{\der \rho} 
		\big\rangle \big] \1_N
	\end{array}\right),
	\\
	& A_1 \equiv  
	\left(\begin{array}{cccc}
		u \kakko{\frac{\der^3 U_k}{\der \sigma^3}} & 
		t \kakko{\frac{\der^2 U_k}{\der \sigma\der \rho}} & 0 
		\\
		t \kakko{\frac{\der^2 U_k}{\der \sigma\der \rho}} &
		u \kakko{\frac{\der^2 U_k}{\der \sigma\der \rho}} & 0
		\\
		0 & 0 & u \kakko{\frac{\der^2 U_k}{\der \sigma\der \rho}} \1_{N-1}
	\end{array}\right), 
	\hspace{-5pt}
\ea
and $A_2$ is a collection of terms at $\calO (u^2, ut, t^2)$. 
Defining $\dert$ as a derivative acting only on the $k$-dependence of $R_k$, 
we obtain 
\ba
	& \der_k \Gamma_{k}\Big|_{\rm kin} 
	\notag
	\\
	& = \frac{1}{2}\dert\Tr \log (\Gamma_k^{(2)}+R_k) \Big|_{\rm kin}
	\notag
	\\
	& = \frac{1}{2}\dert\Tr \log (A_0 + A_1 + A_2 + \dots)\Big|_{\rm kin}
	\notag
	\\
	& = \frac{1}{2}\dert\Tr \Big[\log A_0 + 
	\log \Big\{\1 + A_0^{-1}(A_1+A_2+\dots)\Big\} \Big] \Big|_{\rm kin}
	\notag
	\\
	& = \frac{1\rule{0pt}{15pt}}{2} \dert\Tr \! 
	\left[ - \frac{1}{2}A_0^{-1}A_1 A_0^{-1}A_1 \right] \Big|_{\rm kin} \,,
	\label{eq:4erdfgdf}
\ea
where we have used that $\Tr [ \log A_0 ]$ and $\Tr [ A_0^{-1}A_2]$ 
do not contribute to the kinetic term.  

On the other hand, a direct substitution of \eqref{eq:inhbg} 
into the Ansatz \eqref{eq:ansatz} yields  
\ba
	\der_k \Gamma_{k}\Big|_{\rm kin} 
	& = \frac{1}{2}\int_q \left[ 
	(\der_k Y_k) q^2 t_{q}t_{-q} + (\der_k Z_k) q^2 u_{q}u_{-q} \right] .
	\label{eq:34rtdf}
\ea
Juxtaposing \eqref{eq:4erdfgdf} with \eqref{eq:34rtdf}, we obtain
\begin{widetext}
	\ba
		\der_k Y_k 
		& = - \kakko{\frac{\der^2 U_k}{\der \sigma\der \rho}}^2 
		\lim_{q\to 0}\frac{\der}{\der(q^2)}
		\dert \int_p \frac{1}{ Z_k p^2 + \Rs(p) 
		+ \kakko{ \frac{\der^2 U_k}{\der \sigma^2} }
		} 
		\frac{1}{ Y_k (p+q)^2 + \Rp(p+q) 
		+ \kakko{ \frac{\der U_k}{\der \rho} }
		} \,, 
		\label{eq:YY}
		\\
		\der_k Z_k 
		& = -\frac{1}{2}\kakko{\frac{\der^3 U_k}{\der \sigma^3}}^2  
		\lim_{q\to 0}\frac{\der}{\der(q^2)}\dert 
		\int_p 
		\frac{1}{ Z_k (q+p)^2 + \Rs(q+p) 
		+ \kakko{\frac{\der^2 U_k}{\der \sigma^2}}} 
		\frac{1}{ Z_k p^2 + \Rs(p) + \kakko{\frac{\der^2 U_k}{\der \sigma^2}}}  
		\notag
		\\
		& \quad 
		- \frac{N}{2} \kakko{\frac{\der^2 U_k}{\der \sigma\der \rho}}^2 
		\lim_{q\to 0}\frac{\der}{\der(q^2)}\dert 
		\int_p \frac{1}{ Y_k (q+p)^2 + \Rp(q+p) 
		+ \kakko{\frac{\der U_k}{\der \rho}}} 
		\frac{1}{ Y_k p^2 + \Rp(p) + \kakko{\frac{\der U_k}{\der \rho}}}\,.
		\label{eq:ZZ}
	\ea
\end{widetext}
Finally we evaluate \eqref{eq:YY} and \eqref{eq:ZZ} analytically 
with the help of formulas for threshold functions 
with the optimized regulator in e.g., \cite{Hofling:2002hj,Braun:2008pi}. 
This leads to the relatively simple expressions, \eqref{eq:ephff} 
and \eqref{eq:etff}. 

\bibliographystyle{apsrev4-1}
\bibliography{5dO(N)_arxiv_v1.bbl}
\end{document}